\documentclass[twocolumn]{revtex4}
\usepackage{graphicx}
\usepackage{amsfonts,amsmath,ulem}
\usepackage{float,color,array,multirow}

\begin{document}

\title{Triple-${\bf Q}$ collinear state with compensated ferrimagnetic nature on frustrated kagome lattice}

\author{Kazushi Aoyama$^1$ and Hikaru Kawamura$^2$}

\affiliation{ $^1$Department of Earth and Space Science, Graduate School of Science, The University of Osaka, Osaka 560-0043, Japan \\
$^2$Molecular Photoscience Research Center, Kobe University, Kobe 657-8501, Japan}

\date{\today}

\begin{abstract}
Spin-selective band splitting without net magnetization and spin-orbit couplings serves for a next-generation spin-current generator, and its typical platforms are altermagnets and compensated ferrimagnets as well, where the existence of a crystal asymmetry or nonequivalent sites is essential. Here, we theoretically demonstrate that such a splitting can be realized in a triple-{\bf Q} 12-sublattice state emerging in a $J_3$-dominant kagome antiferromagnet, without the help of the crystal asymmetry. Reflecting the multi-sublattice nature, a local magnetization reveals a fully compensated ferrimagnetic pattern in units of a triangle plaquette, leading to $s$-wave-type spin splittings in magnon and electron bands. This enables an atiferromagnetic spin Seebeck effect at zero field in insulating systems and filling-controlled polarized states in metallic systems, highlighting the potential of frustrated magnets to realize novel spintronics functionalities.
\end{abstract}

\maketitle


\section{Introduction}
Recently, collinear antiferromagnets have attracted renewed interest in the context of altermagnets as a platform for spintronics applications. In the altermagnets \cite{alter_Smejkal_prx_22_overview, alter_Noda_pccp_16, alter_Naka_natcom_19, alter_Hayami_jpsj_19, alter_Yuan_prb_20}, although up and down spins are compensated as in antiferromagnets, the electronic band exhibits a spin-splitting similarly to the ferromagnetic one even in the absence of a spin-orbit coupling. Such an unusual situation also emerges in compensated ferrimagnets \cite{cferri_Groot_prl_95, cferri_Pickett, cferri_Misawa_prl_24,cferri_Liu_prl_25,cferri_Lee_arX_25} where the splitting does not alternate in the momentum space as the up and down spins are cancelled out not by the crystal symmetry \cite{alter_Mazin_review_24}. The common feature of the two is breaking of the effective time-reversal symmetry consisting of the spin-flip and translational/inversion symmetries \cite{alter_Smejkal_prx_22_overview, alter_Hayami_jpsj_19, alter_Naka_natcom_19, alter_Yuan_prb_20, alter_Hernandez_prl_21, alter_Smejkal_prx_22, alter_Yuan_prb_21,cferri_Misawa_prl_24,cferri_Liu_prl_25}, where site asymmetries or anisotropic electronic states around the spin are essential for the broken symmetry. In this work, we theoretically demonstrate that a triple-${\bf Q}$ collinear antiferromagnetic structure emerging on the frustrated kagome lattice breaks the effective time-reversal symmetry without the help of the crystal asymmetry, enabling an isotropic spin-current generation by thermal activation at zero field.  
 
\begin{figure}[t]
\begin{center}
\includegraphics[width=\columnwidth]{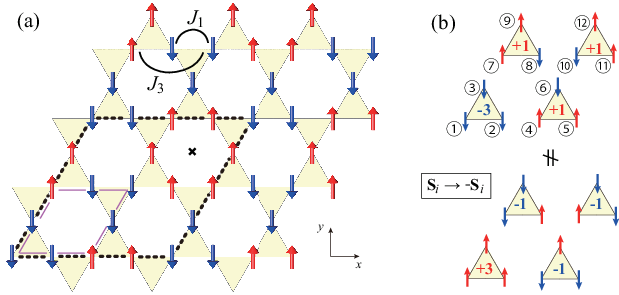}
\caption{The 12 sublattice antiferromagnetic state on the kagome lattice. (a) The real-space spin structure, where the spins are assumed to be oriented in the out-of-plane direction (see the text). The small magenta and large black dotted parallelograms present the crystal and magnetic unit cells, respectively. The cross indicates a rotational center. (b) Zoomed view of the 6 up and 6 down spins within the magnetic unit cell in (a), where these 12 sublattice spins are labeled by \textcircled{\scriptsize 1}-\textcircled{\scriptsize 12}. One can see the ferrimagnetic pattern in units of the triangle plaquette, and nonequivalence between the original (upper) and spin-flipped (lower) spin configurations. 
\label{fig1_1} }
\end{center}
\end{figure}

An antiferromagnetic exchange interaction between neighboring localized spins is usually the source of the collinear antiferromagnetic order. Such a collinear structure often becomes unstable on lattices with geometrical frustration such as triangular and kagome lattices, eventually leading to noncollinear or spin-liquid states. In some frustrated systems, however, thermal and quantum fluctuations favor collinear spin configurations among energetically-degenerate ground-state candidates \cite{fcc_Henly_87, tri_Kawamura_85, tri_Chubkov_91}. One example of such an antiferromagnetic collinear state stabilized by the fluctuation effect is a triple-${\bf Q}$ 12-sublattice state realized in the $J_1$-$J_3$ Heisenberg model on the kagome lattice \cite{RMO-collinear_Grison_prb_20,KagomeSkX_AK_22, KagomeSkX_AK_23} whose real-space spin configuration is shown in Fig. \ref{fig1_1} (a).

The 12 sublattice collinear state consists of $\uparrow\uparrow\downarrow\downarrow$ chains running along the bond directions, and its magnetic unit cell contains four upward triangles each having local magnetization ${\bf m}_{\triangle} ={\bf S}_A+{\bf S}_B+{\bf S}_C$ with three corner spins ${\bf S}_A$, ${\bf S}_B$, and ${\bf S}_C$. As shown in Fig. \ref{fig1_2} (b), ${\bf m}_{\triangle}$ on one triangle is $-3$ and the rest three are $+1$, so that the state can be viewed as a fully compensated ferrimagnet in units of the triangle plaquette. The plaquette ferrimagnetic pattern cannot be superimposed on the spin-flipped one by any spatial symmetry operations, so that the effective time-reversal symmetry is broken, pointing to the occurrence of spin-split bands. A unique aspect of this system is that the asymmetry appears not in the crystal structure or the interaction but in the magnetic structure. 

In this paper, we discuss magnon properties of the 12 sublattice antiferromagnetic state, bearing insulating systems in our mind. We demonstrate that the magnon band exhibits an $s$-wave-type spin splitting, and resultantly, thermally activated magnons can carry net spin current even at zero field. It is also shown that in a metallic system, a similar spin-splitting appears in the band of the conduction electrons coupled to the 12 sublattice spins, pointing to the possibility of filling-dependent spin-polarized states.  

\section{Results}
\subsection{Model} 
We start from the following spin Hamiltonian on the kagome lattice: 
\begin{equation}\label{eq:Hamiltonian_s}
{\cal H}_S  = J_1 \sum_{\langle i,j \rangle} {\bf S}_i\cdot{\bf S}_j + J_3\sum_{\langle \langle  i,j \rangle \rangle } {\bf S}_i\cdot{\bf S}_j ,
\end{equation}
where ${\bf S}_i$ is a classical Heisenberg spin at site $i$, and $J_1>0$ and $J_3>0$ are the nearest neighbor (NN) antiferromagnetic interaction and the third NN one along the bond direction, respectively [see Fig. \ref{fig1_1} (a)]. For relatively strong $J_3$ as realized in the vesignieite BaCu$_3$V$_2$O$_8$(OH)$_2$ \cite{CoplanarOct_Boldrin_prl_18}, a triple-${\bf Q}$ state characterized by the three commensurate ordering vectors ${\bf Q}_1=\frac{\pi}{2}(-1,-\frac{1}{\sqrt{3}})$, ${\bf Q}_2=\frac{\pi}{2}(1,-\frac{1}{\sqrt{3}})$, and ${\bf Q}_3=\frac{\pi}{2}(0,\frac{2}{\sqrt{3}})$ is favored. The ${\bf Q}_1$, ${\bf Q}_2$, and ${\bf Q}_3$ are associated with the left-, right-, and top-corner spins on the upward triangles ${\bf S}_A$ , ${\bf S}_B$, and ${\bf S}_C$, respectively, each constituting an $\uparrow\downarrow\uparrow\downarrow$ collinear configuration. The relative angles, i.e, superposition angles,  among ${\bf S}_A$, ${\bf S}_B$, and ${\bf S}_C$ cannot be determined from the ground-state energy. The ground-state degeneracy with respect to the relative angles, however, is lifted by the thermal fluctuation which, in the present case, favors the collinear configuration shown in Fig. \ref{fig1_1}, the 12 sublattice state with the plaquette ferrimagnetic structure.  

Since the collinear state is already favored at the isotropic Heisenberg level, easy-plane and easy-axis anisotropies typically present in real materials merely orient the collinear axis in the easy-plane and easy-axis directions, respectively, as it has been confirmed in Monte Carlo simulations \cite{KagomeSkX_AK_23}. In the easy-plane case, the collinear state is selected by the thermal fluctuation similarly to the Heisenberg case, whereas in the easy-axis case, it is firmly stabilized by the anisotropy energy. Since the triple-${\bf Q}$ collinear state of our interest is robust against these single-ion anisotropies, in this paper, we consider the isotropic Heisenberg model as the minimal model.
Note that in the two-dimensional Heisenberg-spin model, a long-range magnetic order is not allowed at any finite temperature, and the transition into the collinear state can be characterized by the spatial distribution of ${\bf m}_\triangle$ or a bond variable \cite{KagomeSkX_AK_22,RMO-collinear_Grison_prb_20}. Of course, the long-range or quasi-long-range magnetic order is guaranteed in the presence even of an infinitesimally small magnetic anisotropy \cite{KagomeSkX_AK_23}. 

\subsection{Insulating system} 
The 12 sublattice collinear spins $\overline{\bf S}_\mu$ with sublattice indices $\mu=1,2, 3, \cdots ,12$ defined in Fig. \ref{fig1_1} (b) can be expressed as $\overline{\bf S}_\mu = S m_\mu  \, \hat{z}$ with $S$ being the size of the spin and ${\bf m}=(-1,-1,-1,1,1,-1,1,-1,1,-1,1,1)$. Since in the present isotropic Heisenberg model, the collinear axis can be taken in any direction, in Fig. \ref{fig1_1}, it is assumed that the spins are oriented in the out-of-plane direction for ease of understanding. As discussed in the introduction, this antiferromagnetic structure breaks the effective time-reversal symmetry with the Hamiltonian (\ref{eq:Hamiltonian_s}) itself being spatially symmetric. In this work, we first discuss the magnon properties bearing insulating systems in our mind.

In the present classical spin system, the spin-wave or magnon dispersion can be derived by using the Holstein-Primakoff transformation which involves the creation (annihilation) operator of the magnon on sublattice $\mu$ $\hat{a}_{\bf q}^{\mu \dagger}$ ($\hat{a}_{\bf q}^{\mu}$). By introducing $\mbox{\boldmath $\Phi$}_{B{\bf q}}^\dagger =(\hat{a}_{\bf q}^{1\, \dagger},  \hat{a}_{\bf q}^{2\, \dagger}, \cdots , \hat{a}_{\bf q}^{N \, \dagger},  \hat{a}_{-{\bf q}}^{1}, \hat{a}_{-{\bf q}}^{2}, \cdots , \hat{a}_{-{\bf q}}^{N} )$ with the total number of sublattices $N=12$, we can rewrite the spin Hamiltonian (\ref{eq:Hamiltonian_s}) as 
\begin{eqnarray}\label{eq:Hamiltonian_magnon}
&& {\cal H}_S\simeq \frac{S}{2} \sum_{\bf q} \mbox{\boldmath $\Phi$}_{B{\bf q}}^\dagger  H_{B{\bf q}} \mbox{\boldmath $\Phi$}_{B{\bf q}}, \quad H_{B{\bf q}}= \left(\begin{array}{cc}
A_{\bf q} & B_{\bf q} \nonumber\\
B_{{\bf q}} & A_{{\bf q}} \nonumber
\end{array} \right), \nonumber\\
&& A_{\bf q} = 4J_3 I_{N\times N}+\left( \begin{array}{cccc}
D_0 & 0 & 0 & 0 \\
0 & D_1 & D_2^\ast & D_3^\ast \\
0 & D_2^\ast & D_2 & D_1^\ast \\
0 & D_3^\ast & D_1^\ast & D_3 
\end{array}\right), \nonumber\\
&& B_{\bf q} = \left( \begin{array}{cccc}
0 & C_1+D_1^\ast & C_3+D_3^\ast & C_2+D_2^\ast \\
C_1+D_1^\ast & D_2+D_3 & C_2 & C_3 \\
C_3+D_2^\ast & C_2 & D_1+D_2 & C_1 \\
C_2+D_2^\ast & C_3 & C_1 & D_1+D_3 
\end{array}\right),  \nonumber\\
&& D_l = J_1 \left( \begin{array}{ccc}
0& G_{l1,{\bf q}} &  G_{l3,{\bf q}}^\ast  \nonumber\\
G_{l1,{\bf q}}^\ast & 0 & G_{l2,{\bf q}} \nonumber\\
G_{l3,{\bf q}} & G_{l2,{\bf q}}^\ast & 0  \nonumber
\end{array}\right), \nonumber\\
&&C_l = J_3 \left( \begin{array}{ccc}
T_{l1,{\bf q}} + T_{l3,{\bf q}} &0 &  0 \nonumber\\
0 & T_{l1,{\bf q}} + T_{l2,{\bf q}} & 0 \nonumber\\
0 &0 & T_{l2,{\bf q}} + T_{l3,{\bf q}}  \nonumber
\end{array}\right), \nonumber\\
&& G_{lm, {\bf q}} =  s_{lm} \, e^{i{\bf q}\cdot \hat{e}_m}, \quad T_{lm,{\bf q}} = s_{lm} \, 2 \cos(2{\bf q}\cdot \hat{e}_m), 
\end{eqnarray} 
where $s_{lm} = \delta_{l m}+\delta_{l 0}$ and $\hat{e}_1=(1,0)$, $\hat{e}_2=(-\frac{1}{2}, \frac{\sqrt{3}}{2})$, and $\hat{e}_3=(-\frac{1}{2},-\frac{\sqrt{3}}{2})$ are the real-space unit vectors along the bond directions.
The magnon Hamiltonian (\ref{eq:Hamiltonian_magnon}) can be diagonalized with the use of a para-unitary matrix $T_{\bf q}$ as \cite{LinearRes_Matsumoto_prb_14} 
\begin{equation}
T_{\bf q}^\dagger H_{B{\bf q}} T_{\bf q}=\left( \begin{array}{cc}
E_{\bf q} & 0 \\
0 & E_{-{\bf q}} 
\end{array} \right) , 
E_{\bf q} = \left(\begin{array}{ccc}
\epsilon_{1, {\bf q}} & & 0 \\
 & \ddots & \\
 0 && \epsilon_{N, {\bf q}} 
\end{array} \right) . \nonumber
\end{equation}
Since $\sigma_3 H_{B{\bf q}} T_{\bf q}= T_{\bf q} \left( \begin{array}{cc}
E_{\bf q} & 0 \\
0 & -E_{-{\bf q}} 
\end{array} \right)$ with $\sigma_3 = I_{N\times N} \otimes (-I)_{N\times N}$ is satisfied, $\epsilon_{n, {\bf q}}$ and $T_{\bf q}$ can be obtained by solving the eigen value problem for $\sigma_3 H_{B{\bf q}}$. 

\begin{figure}[t]
\begin{center}
\includegraphics[width=\columnwidth]{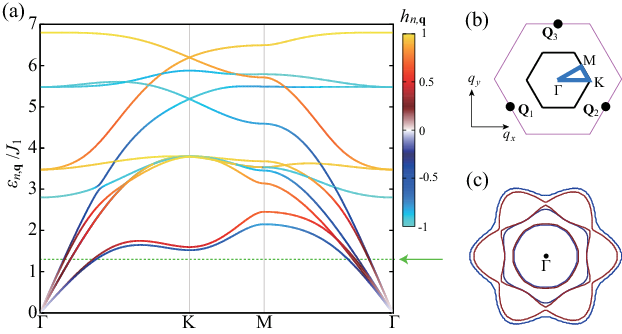}
\caption{The magnon dispersion $\epsilon_{n,{\bf q}}$ in the 12 sublattice state for $J_3/J_1=1.2$. (a) $\epsilon_{n,{\bf q}}$ along the path connecting high-symmetry points in the momentum space shown in (b), where the color represents the effective spin polarization $h_{n,{\bf q}}$. In (b), the inner black and outer magenta hexagons represent the Brillouin zones associated with the magnetic unit cell [the large black dotted parallelogram in Fig. \ref{fig1_1} (a)] and the crystal unit cell [the small magenta parallelogram in Fig. \ref{fig1_1} (a)], respectively. (c) The contour plot at $\varepsilon_{n,{\bf q}}/J_1=1.3$. \label{fig1_2} }
\end{center}
\end{figure}

Figure \ref{fig1_2} (a) shows the magnon dispersion; there are 12 branches some of which are partly degenerate. To characterize the magnon band $\epsilon_{n, {\bf q}}$, we introduce the quantity $h_{n,{\bf q}} = \sum_{\mu=1}^{N} (-m_\mu) ( |T_{{\bf q},\mu n}|^2 + |T_{{\bf q},\mu n+N}|^2)$. Noting that the fluctuation of the net magnetization around zero is given by 
$\sum_{\bf q}\sum_{\mu=1}^{N}m_{\mu}(S-\langle \hat{a}^{\mu \dagger}_{\bf q}\hat{a}^{\mu}_{\bf q} \rangle)=\sum_{\bf q}\sum_{n =1}^{N} f_B(\epsilon_{n,{\bf q}}) h_{n,{\bf q}} + \delta M_0$ with the Bose distribution function $f_B(x)=\frac{1}{e^{x/T}-1}$ and zero-point fluctuation contribution $\delta M_0 = \sum_{\bf q} \sum_{n,\mu=1}^N (-m_\mu)  |T_{{\bf q},\mu n+N}|^2$, we may interpret $h_{n,{\bf q}}$ as the effective spin polarization of the $n$th magnon band. 
 Since in the conventional 2-sublattice case on the square lattice where $T_{\bf q}$ corresponds to the Bogoliubov transformation matrix 
$\small\left( \begin{array}{cc}
 \cosh \theta_{\bf q} \, I_{2\times 2}& -\sinh \theta_{\bf q} \, \sigma_x \\
 -\sinh\theta_{\bf q} \, \sigma_x & \cosh\theta_{\bf q} \, I_{2\times 2}
\end{array} \right)$ with the Pauli matrix $\sigma_x$, $h_{n,\bf q}$ can be calculated as +1 and -1 for left-handed and right-handed polarized states, respectively, $h_{n,{\bf q}}$ could also be understood as a generalized form of the magnon chirality \cite{alter_magnon_Smejkal_prl_23, alter_magnon_Liu_prl_24,MagnonPol_Nambu_prl_20, MagnonPol_Kawamoto_apl_24}.

One can see from Fig. \ref{fig1_2} (a) that the magnon band exhibit a spin splitting, and the lowest two spin-split bands, which should be relevant to the magnon transport, do not intersect over the whole momentum space, as it can clearly be seen from the outer red and blue closed loops in the contour plot of $\epsilon_{n, {\bf q}}$ shown in Fig. \ref{fig1_2} (c). Such a band structure is similar to electronic bands subject to the Zeeman splitting rather than the altermagnetic one whose spin polarization alternates in the momentum space \cite{alter_Smejkal_prx_22_overview, alter_Hayami_jpsj_19, alter_Naka_natcom_19, alter_Yuan_prb_20, alter_Hernandez_prl_21, alter_Smejkal_prx_22, alter_Yuan_prb_21}. The 6-fold symmetry of the non-alternating magnon band originates from the fact that the spin configuration possesses the rotational symmetry around the center of the hexagon with $\uparrow$ spins on its six corners [see the cross in Fig. \ref{fig1_1} (a)]. 
From the viewpoint of symmetry, the kagome lattices without and with the nonequivalent up and down sites possess the same $D_{6h}$ point group symmetry. The difference between the two consists in the translational symmetry; the kagome lattice itself is invariant under the translation by a primitive vector [side vectors of the small magenta parallelogram in Fig. \ref{fig1_1} (a)], whereas the kagome lattice with the up and down sties is not invariant as these sites are distinguished. Such a noninvariance cannot be recovered by the additional spin flip operation, namely, the effective time-reversal symmetry is broken. Thus, the spin splitting occurs.
Reflecting the $s$-wave-type spin splitting, the up-spin magnon always has the larger velocity (the steeper slope near the $\Gamma$ point) than the down-spin magnon, so that the up-spin component of the magnon may govern the low-energy magnon transport, pointing to ferromagnetic features in the magnon transport. 

Although our focus is on the excitation from the antiferromagnetic ground state of the classical spins, we briefly comment on to the quantum correction to the ground state. 
First, the leading order correction to the energy, i.e., the linear contribution in the $1/S$ expansion, assists the occurrence of the collinear state \cite{RMO-collinear_Grison_prb_20}, suggesting that both thermal and quantum fluctuations favor the collinear state. Although in the spin split case, nonlinear contributions may yield nontrivial effects \cite{alter_qfluc_Eto_prb_25,alter_qfluc_Cichutek_prb_25}, the ground-state stability would be basically dominated by the leading order contribution as long as the $1/S$ expansion converges for large $S$. Second, the zero-point quantum-fluctuation contribution to the magnetization can numerically be evaluated as $\delta M_0 \simeq 0.002$ for the parameter used here and thus, the compensation is imperfect at the quantum mechanical level. Although the association between the magnon band and $\delta M_0$ is not clear, it is certain that the $s$-wave-type spin splitting indicates the emergence of possible ferromagnetic properties. 

To see how the spin splitting affects the magnon transport, we examine the spin current driven by a thermal gradient. The spin current operator can be derived from the conservation law for the magnetization as ${\bf J}_s^z = \sum_i {\bf r}_i \, \frac{d}{dt}S^z_i = i \sum_i {\bf r}_i \, [H_S, S^z_i]$ \cite{book_Mahan, trans-sq_AK_prb_19, trans-tri_AK_prl_20}. With the use of the velocity matrix $v_{{\bf q},\mu} = \frac{\partial}{\partial q_\mu} H_{B{\bf q}}$ and the diagonal matrix $(I_m)_{ij} =(\sigma_3)_{ij} (-m)_j$, it can be expressed as $J^z_{s,\mu}=\sum_{\bf q} \mbox{\boldmath $\Phi$}_{B{\bf q}}^\dagger  V^s_{{\bf q},\mu} \mbox{\boldmath $\Phi$}_{B{\bf q}} $ with $V^s_{{\bf q},\mu}= \frac{S}{4} (v_{{\bf q},\mu} I_m + I_m v_{{\bf q},\mu})$. Also, the thermal current is given by $J_{th,\mu}=  \sum_{\bf q} \mbox{\boldmath $\Phi$}_{B{\bf q}}^\dagger V^{th}_{{\bf q},\mu}  \mbox{\boldmath$\Phi$}_{B{\bf q}}$ with $V^{th}_{{\bf q},\mu} = \frac{S^2}{4}( v_{{\bf q},\mu} \sigma_3 H_{B{\bf q}}+ H_{B{\bf q}} \sigma_3 v_{{\bf q},\mu}) $ \cite{LinearRes_Matsumoto_prb_14}. 
In the linear response theory, the spin current generated by the thermal gradient is characterized by the associated conductivity $\chi^{\rm SC}_{\mu \nu} = \frac{i}{T} \frac{d Q_{\mu\nu}(\omega + i 0 )}{d \omega}|_{\omega=0}$ with $Q_{\mu \nu}(i\omega_m) = - \frac{1}{V}\int_0^{1/T}\langle T_\tau J^z_{s,\mu} (\tau) J_{th, \nu}(0) \rangle e^{i\omega_m \tau} d\tau$. By using a relaxation-time approximation where the inverse magnon scattering time $\gamma=1/\tau_m$ is assumed to be a constant for simplicity, we can evaluate $\chi^{\rm SC}_{\mu\nu}$ as (for details, see Ref. \cite{supple})
\begin{equation}\label{eq:response}
\chi^{\rm SC}_{\mu \nu} =  -\frac{1}{VT}\frac{1}{\gamma} \sum_{\bf q} \sum_{n=1}^{2N}  f^\prime_B(\epsilon_{n,{\bf q}})(T_{\bf q}^\dagger V^s_{{\bf q},\mu} T_{\bf q} )_{nn} (T_{\bf q}^\dagger V^{th}_{{\bf q},\nu}T_{\bf q})_{nn} ,
\end{equation}
with $f^\prime_B(x) = \frac{d}{dx}f_B(x)$. In the same manner, the thermal conductivity $\kappa_{\mu \nu} $ can be calculated by simply replacing $V^s_{{\bf k},\mu}$ with $V^{th}_{{\bf k},\mu}$ in Eq. (\ref{eq:response}).  

\begin{figure}[t]
\begin{center}
\includegraphics[width=.9\columnwidth]{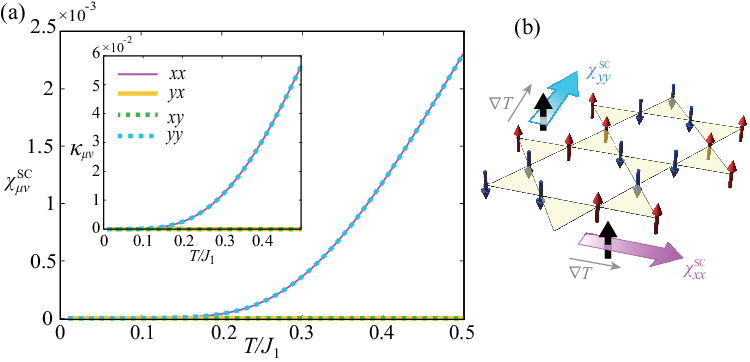}
\caption{ (a) The temperature dependence of the spin current conductivity under a temperature gradient $\chi^{\rm SC}_{\mu\nu}$ for the same parameter as that for Fig. \ref{fig1_2}, where magent, orange, green, and cyan curves denote the $xx$, $yx$, $xy$, and $yy$ components of $\chi^{\rm SC}_{\mu\nu}$, respectively. The inset shows the associated thermal conductivity $\kappa_{\mu \nu}$, where the color notations are the same as those in the main panel. $\chi^{\rm SC}_{\mu\nu}$ and $\kappa_{\mu\nu}$ are normalized by $S^3J_1/\gamma$ and $S^4J_1^2/\gamma$, respectively. (b) An image of the system, where the thermal gradient $\nabla T$ is applied in in-plane directions.\label{fig2} }
\end{center}
\end{figure}

Figure \ref{fig2} (a) shows the temperature  dependence of $\chi^{\rm SC}_{\mu \nu}$ (main panel) and $\kappa_{\mu\nu}$ for reference (inset). For both the spin and thermal currents, the longitudinal responses to the temperature gradient $\chi^{\rm SC}_{\mu \mu}$ and $\kappa_{\mu\mu}$ are nonzero at finite temperatures, whereas the transverse responses are absent. Also, the $xx$ and $yy$ components of the longitudinal responses [see magenta and cyan symbols in Fig. \ref{fig2} (a)] are almost identical to each other. Since the temperature dependence of $\chi^{\rm SC}_{\mu \mu}$ is qualitatively the same as that of $\kappa_{\mu\mu}$, it is indicated that in the 12 sublattice antiferromagnetic state, thermally activated magnons contribute the spin current similarly to ferromagnetic magnons. This is in sharp contrast to the conventional antiferromagnets where the up and down spin currents are completely canceled out at zero field and thus, $\chi^{\rm SC}_{\mu \mu}= 0$. This unique feature of the present system, i.e., the emergence of the lateral spin Seebeck effect without net magnetization as in the case of the compensated ferrimagnets \cite{cferi_Seebeck_Maekawa_prb_13}, originates from the difference in the velocity between the up-spin and down-spin magnons, which should be more remarkable as the associated spin splitting becomes larger. 

\subsection{Metallic system}
\begin{figure}[t]
\begin{center}
\includegraphics[width=\columnwidth]{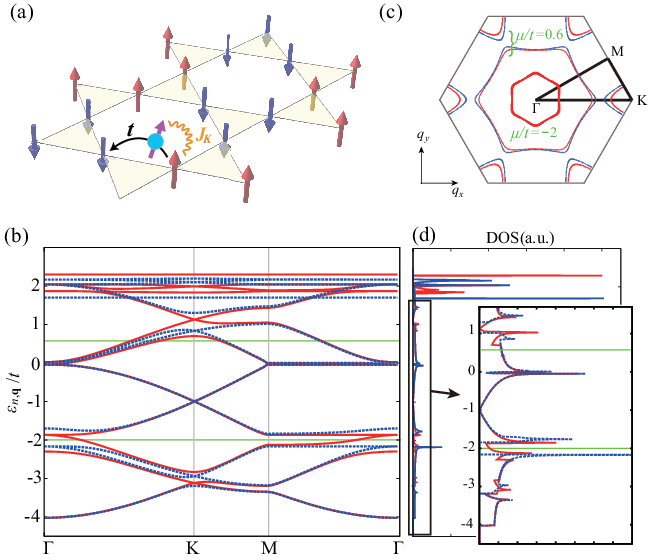}
\caption{The band structure of the electrons coupled to the 12 sublattice antiferromagnetic collinear spins for $J_K/t=0.3$. (a) The schematic picture of the Kondo lattice model. (b) The spin-$\uparrow$ (red) and spin-$\downarrow$ (blue) electron dispersions along the path connecting high-symmetry points shown in (c) where the contour plots at $\varepsilon_{n,{\bf q}}/t=0.6$ (outer part) and $-2$ (inner part) are shown. (d) The associated density of states (DOS), where the inset shows a zoomed view of the region enclosed by a box.  \label{fig3} }
\end{center}
\end{figure}

Now that the magnon properties are understood, we shall next discuss how the broken effective time-reversal symmetry in the spin sector is reflected in electron bands. We consider a Kondo-lattice model, ${\cal H}_{e} = -t\sum_{\langle i,j\rangle}\sum_{s=\uparrow \downarrow} \big( \hat{c}_{i s}^\dagger \hat{c}_{j s}+\hat{c}_{i s}\hat{c}_{j s}^\dagger \big) + J_K \sum_{i, \alpha,\beta} {\bf S}_i \cdot \mbox{\boldmath $\sigma$}_{\alpha \beta} \hat{c}_{i\alpha}^\dagger \hat{c}_{i\beta}$, in which electrons move on the kagome lattice with the NN hopping $t$, interacting with the classical localized spin ${\bf S}_i$ via the Kondo coupling $J_K$ [see Fig. \ref{fig3} (a))]. For simplicity, we fix ${\bf S}_i$ to be the 12 sublattice configuration, assuming that it is determined by the exchange interactions without being affected by the Kondo coupling. Then, the problem is reduced to diagonalizing the following electronic Hamiltonian ${\cal H}_{e}$    
 \begin{eqnarray}\label{eq:Hamiltonian_electron}
&& {\cal H}_{e}= \sum_{\bf q} \mbox{\boldmath $\Phi$}_{F{\bf q}}^\dagger  H_{F{\bf q}} \mbox{\boldmath $\Phi$}_{F{\bf q}}, \quad H_{F{\bf q}}= \left(\begin{array}{cc}
M_{\bf q} & 0 \nonumber\\
0 & M_{{\bf q}} \nonumber
\end{array} \right) + J_K I_m, \nonumber\\
&& M_{\bf q} = \left( \begin{array}{cccc}
D^\prime_0 & D^{\prime \ast}_1 & D^{\prime \ast}_3 & D^{\prime \ast}_2 \\
D^{\prime \ast}_1 & D^\prime_0 & D^{\prime \ast}_2 & D^{\prime \ast}_3 \\
D^{\prime \ast}_3 & D^{\prime \ast}_2 & D^\prime_0 & D^{\prime \ast}_1 \\
D^{\prime \ast}_2 & D^{\prime \ast}_3 & D^{\prime \ast}_1 & D^\prime_0 
\end{array}\right), \quad D^\prime_l  = (-t/J_1) D_l
\end{eqnarray}  
with $\mbox{\boldmath $\Phi$}_{F{\bf q}}^\dagger =(\hat{c}_{{\bf q}\uparrow}^{1\, \dagger},  \hat{c}_{{\bf q}\uparrow}^{2\, \dagger}, \cdots , \hat{c}_{{\bf q}\uparrow}^{N \, \dagger},  \hat{c}_{{\bf q}\downarrow}^{1}, \hat{c}_{{\bf q}\downarrow}^{2}, \cdots , \hat{c}_{{\bf q}\downarrow}^{N} )$, where $\hat{c}_{{\bf q}s}^{\mu \dagger}$ ($\hat{c}_{{\bf q}s}^{\mu}$) is the creation (annihilation) operator of the spin-$s$ electron on sublattice $\mu$. In Eq. (\ref{eq:Hamiltonian_electron}), the matrix $(I_m)_{ij} =(\sigma_3)_{ij} (-m)_j$ possesses the information of the localized spins. 

Figure \ref{fig3} (b) shows the spin-$\uparrow$ (red solid curve) and spin-$\downarrow$ (blue dotted curve) electronic bands for $J_K/t=0.3$, and examples of associated Fermi surfaces are shown in Fig. \ref{fig3} (c). The electron band exhibits a spin splitting as expected from the symmetry consideration, and the spin polarization does not alternate in the momentum space [see the outer part of Fig. \ref{fig3} (c)] similarly to the polarization of the magnon band. Reflecting the non-alternating nature, the Fermi surface in the present antiferromagnetic state looks like a Zeeman split one. Actually, for the chemical potential value of $\mu/t=-2$ [see the inner part of Fig. \ref{fig3} (c)], only the spin-$\uparrow$ Fermi surface appears, pointing to the emergence of a spin-polarized metallic state. 
The origin of such a spin-polarized state consists in nonequivalence in the density of states (DOS) of the $\uparrow$ and $\downarrow$ bands. As shown in Fig. \ref{fig3} (d), the spin splitting is reflected as the difference in the DOS, which can clearly  be seen, for example, in high-energy sharp peaks stemming from the flat-band nature of the kagome lattice. Thus, the electronic state can be positively or negatively spin-polarized depending on the position of the chemical potential or the electron filling. The nonequivalent DOSs of this kind is characteristic of compensated ferrimagnets \cite{cferri_Misawa_prl_24,cferri_Liu_prl_25}. In the present 12 sublattice antiferromagnetic state, the ferrimagnetic spatial pattern of the plaquette magnetization ${\bf m}_\triangle$ plays a role similar to nonequivalent sites in the compensated ferrimagnets. Thus, although the size of the splitting depends on the value of $J_K$, it is certain that the nonequivalent DOS should appear as long as $J_K$ is nonzero.

\section{Discussion}
In this paper, we have theoretically investigated the $J_1$-$J_3$ kagome antiferromagnets, putting an emphasis on band structures in the triple-{\bf Q} state with the 12 sublattice collinear configuration. 
Although the complexity originating from the triple-{\bf Q} nature often yields non-collinear spin textures such as skyrmion crystals, in the present collinear case, it works as an internal asymmetry, i.e., the underlying cause that breaks the effective time-reversal symmetry defined by the combination of the spin-flip operation and lattice translation. Specifically, reflecting the triple-{\bf Q} multi-sublattice nature, the local magnetization defined on the triangle plaquette reveals a fully compensated ferrimagnetic pattern. As a consequence, magnon and electron bands exhibit $s$-wave-type spin splittings. In insulators, since the up- and down-spin magnons are nonequivalent over the whole momentum space due to the spin splitting, a thermal gradient can drive a net magnon spin current even at zero field. In a metallic system, the splitting leads to nonequivalent DOSs for the up- and down-spin electronic bands, suggesting that the electronic state can be polarized either positively or negatively by tuning the electron filling.
 
Experimentally, the occurrence of the triple-{\bf Q} 12-sublattice state has been reported in the $J_1$-$J_3$ kagome antiferromagnet BaCu$_3$V$_2$O$_8$(OH)$_2$\cite{CoplanarOct_Boldrin_prl_18}, but its spin structure is not theoretically-expected collinear but rather coplanar. Nevertheless, it would be useful to learn how such an unusual $J_3$ dominant situation can be realized. In this kagome antiferromagnet, each magnetic Cu$^{2+}$ site possesses orbital degrees of freedom in the $e_g$ level, the degeneracy of which is usually lifted by the Jahn-Teller (JT) effect, i.e., a spontaneous distortion of the O octahedron surrounding Cu$^{2+}$. In Refs. \cite{CoplanarOct_Boldrin_prl_18,Sr-vesignieite_Boldrin_jmc_15, Ba-vesignieite_Boldrin_jmc_16}, it is proposed that in this class of magnets having noequivalent Cu$^{2+}$ sites, not only static but also dynamical JT effects are relevant and resultant tiny but complicated octahedron distortions lead to the emergence of two types of superexchange paths Cu-O-Cu. For one path, the Cu-O-Cu angle is reported to be $\lesssim 90^{\circ}$ and for the other path, it is $\sim 100^{\circ}$, so that according to the Goodenough-Kanamori rules, the superexchange interactions for the former and the latter are ferromagnetic and antiferromagnetic, respectively, canceling each other. As a result, the net NN interaction $J_1$ is suppressed and the further neighbor interaction $J_3$ becomes dominant. Thus, in general, kagome antiferromagnets possessing several nonequivalent NN superexchange paths for whatever reasons could be $J_3$-dominant candidates. Furthermore, if additional easy-axis anisotropy exists, it enhances the stability of the collinear triple-${\bf Q}$ state by counteracting the Dzyaloshinskii-Moriya interactions inherent to kagome systems that tend to cant the spins \cite{Sr-vesignieite_DM_Verrier_prb_20}.

Although direct candidate materials are not currently available, recent progresses in the artificial fabrication of two-dimensional kagome magnets \cite{Artificial-kagome_Zhou_natcom_24} suggest that the compensated plaquette ferrimagnetic state could be realized in the future by combining the above material insights and fabrication techniques. 
In the realization, orbital degrees of freedom, which can also be the origin of the spin splitting \cite{orbital_alter_Mook_prl_24}, may come into play, making the spin ordering mechanism rather complicated. Nevertheless, considering that the compensated ferrimagnetic state cannot be superimposed on the spin flipped state by any spatial symmetry operation, the $s$-wave nature of the spin splitting would remain unchanged as long as additional orbital degrees of freedom, which may effectively reduce the spatial symmetry, work only as secondary effects. Once multi-sublattice collinear antiferromagnetic states of this kind are realized in kagome and other frustrated systems, thermal activation of the spin current, i.e., the field-free spin Seebeck effect, will become possible, which also enriches the potential of the flatband engineering \cite{Flat_Mizoguchi_prb_19}. The mechanism presented here thus provides an additional guiding principle for the search for magnets with non-relativistic spin-splitting as a platform for novel spintronics functionalities. 
 
\begin{acknowledgments}
The authors thank Y. Niimi, H. Adachi, J. Ieda, Y. Araki, T. Nomoto, and T. Misawa for useful discussions. This work is supported by JSPS KAKENHI Grant No. JP23H00257 and JP24K00572.
\end{acknowledgments}

\pagebreak

\hspace{2cm}
\pagebreak

\onecolumngrid
\hspace{2cm}
\begin{center}
\textbf{\large Supplemental Material for ''Triple-${\bf Q}$ collinear state with compensated ferrimagnetic nature on frustrated kagome lattice''}\\[.2cm]
\end{center}

In this supplemental material, we derive the expressions of the conductivities $\chi^{\rm SC}_{\mu \nu}$ and $\kappa_{\mu \nu}$ in the relaxation-time approximation. We start from the Bogoliubov Hamiltonian for the magnon which can be diagonalized as
\begin{equation}
 {\cal H}_S = \frac{S}{2} \sum_{\bf q} \mbox{\boldmath $\Phi$}_{B{\bf q}}^\dagger (T^\dagger_{\bf q})^{-1} T^\dagger_{\bf q}H_{B{\bf q}} T_{\bf q} T_{\bf q}^{-1} \mbox{\boldmath $\Phi$}_{B{\bf q}} = \frac{S}{2} \sum_{\bf q}  \mbox{\boldmath $\varphi$}_{{\bf q}}^\dagger \left( \begin{array}{cc}
 E_{\bf q} & 0 \\
 0 & E_{-{\bf q}} 
 \end{array} \right) \mbox{\boldmath $\varphi$}_{{\bf q}} . \nonumber
\end{equation}
With the use of the new basis $\mbox{\boldmath $\varphi$}^\dagger_{{\bf q}} = \mbox{\boldmath $\Phi$}_{B{\bf q}}^\dagger (T^\dagger_{\bf q})^{-1} 
= (\hat{\alpha}_{\bf q}^{1\, \dagger},  \hat{\alpha}_{\bf q}^{2\, \dagger}, \cdots , \hat{\alpha}_{\bf q}^{N \, \dagger},  \hat{\alpha}_{-{\bf q}}^{1}, \hat{\alpha}_{-{\bf q}}^{2}, \cdots , \hat{\alpha}_{-{\bf q}}^{N} )$, the spin and thermal current can be  expressed as
 \begin{eqnarray}
&&J^z_{s,\mu}  = \sum_{\bf q} \mbox{\boldmath $\varphi$}_{B{\bf q}}^\dagger  \big( T^\dagger_{\bf q}  V^s_{{\bf q},\mu} T_{\bf q} \big) \mbox{\boldmath $\varphi$}_{B{\bf q}} \equiv \sum_{\bf q} \mbox{\boldmath $\varphi$}_{B{\bf q}}^\dagger   \tilde{V}^s_{{\bf q},\mu}  \mbox{\boldmath $\varphi$}_{B{\bf q}},  \nonumber\\
&& J_{th,\mu} = \sum_{\bf q} \mbox{\boldmath $\varphi$}_{B{\bf q}}^\dagger \big( T^\dagger_{\bf q}  V^{th}_{{\bf q},\mu} T_{\bf q} \big)  \mbox{\boldmath$\varphi$}_{B{\bf q}} \equiv \sum_{\bf q} \mbox{\boldmath $\varphi$}_{B{\bf q}}^\dagger  \tilde{V}^{th}_{{\bf q},\mu}  \mbox{\boldmath$\varphi$}_{B{\bf q}}, \nonumber
\end{eqnarray} 
where as introduced in the main text, $V^s_{{\bf q},\mu}=\frac{S}{4}(v_{{\bf q},\mu} I_m + I_m v_{{\bf q},\mu})$ and $V^{th}_{{\bf q},\mu}=\frac{S^2}{4}(v_{{\bf q},\mu} \sigma_3 H_{B{\bf q}} + H_{B{\bf q}} \sigma_3 v_{{\bf q},\mu} )$.
Then, we have 
\begin{equation}
\langle T_\tau J^z_{s,\mu} (\tau) J_{th, \nu}(0) \rangle  = \sum_{{\bf k},{\bf k}^\prime}  \sum_{m,n,m^\prime,n^\prime=1}^{2N}\langle T_\tau \varphi^\dagger_{{\bf k},m} (\tau)\varphi_{{\bf k},n} (\tau)  \varphi^\dagger_{{\bf k}^\prime,m^\prime} (0) \varphi_{{\bf k}^\prime,n^\prime} (0) \rangle \big( \tilde{V}^s_{{\bf k},\mu} \big)_{mn} \big( \tilde{V}^{th}_{{\bf k}^\prime,\nu} \big)_{m^\prime n^\prime} \nonumber
\end{equation}
whose thermal-average part $\langle ... \rangle$ is reduced to
\begin{equation}\label{eq:wick}
 \langle T_\tau \varphi^\dagger_{{\bf k},m} (\tau)\varphi_{{\bf k},n} (\tau)  \varphi^\dagger_{{\bf k}^\prime,m^\prime} (0) \varphi_{{\bf k}^\prime,n^\prime} (0) \rangle = \langle T_\tau \varphi^\dagger_{{\bf k},m} (\tau) \varphi_{{\bf k}^\prime,n^\prime} (0) \rangle \langle \varphi_{{\bf k},n} (\tau)  \varphi^\dagger_{{\bf k}^\prime,m^\prime} (0) \rangle +  \langle T_\tau \varphi^\dagger_{{\bf k},m} (\tau)  \varphi^\dagger_{{\bf k}^\prime,m^\prime} (0)\rangle \langle T_\tau \varphi_{{\bf k},n} (\tau) \varphi_{{\bf k}^\prime,n^\prime} (0) \rangle.  \nonumber
 \end{equation}
The final expression in Eq. (\ref{eq:wick}) can further be rewritten with the use of the magnon Green's function ${\cal G}^n_{\bf k}(\tau) = -\langle T_\tau \hat{\alpha}^n_{{\bf k}} (\tau)  \hat{\alpha}^{n \dagger}_{{\bf k}} (0) \rangle$. For example, the first term in Eq. (\ref{eq:wick}) is given by
\begin{eqnarray}
&& \delta_{{\bf k}{\bf k}^\prime} \delta_{m n^\prime} \delta_{n m^\prime} \left\{ \begin{array}{l}
 \langle T_\tau \hat{\alpha}^{m \dagger}_{{\bf k}} (\tau) \hat{\alpha}^m_{{\bf k}} (0) \rangle \, \, \,  (1\leq m \leq N) \\
 \langle T_\tau \hat{\alpha}^{m}_{-{\bf k}} (\tau) \hat{\alpha}^{m \dagger}_{-{\bf k}} (0) \rangle \, (N+1\leq m \leq 2N) 
\end{array} \right\} \left\{ \begin{array}{l}
 \langle T_\tau \hat{\alpha}^n_{{\bf k}} (\tau)  \hat{\alpha}^{n \dagger}_{{\bf k}} (0) \rangle  \quad (1\leq n \leq N) \\
 \langle T_\tau \hat{\alpha}^{n \dagger}_{-{\bf k}} (\tau)  \hat{\alpha}^{n}_{-{\bf k}} (0) \rangle \, (N+1 \leq n \leq 2N) 
 \end{array}\right\} \nonumber\\
&&=   \delta_{{\bf k}{\bf k}^\prime} \delta_{m n^\prime} \delta_{n m^\prime} \big( \delta_{m l} \, {\cal G}^l_{\bf k}(-\tau) + \delta_{m l+N} \, {\cal G}^{l}_{-{\bf k}}(\tau) \rangle \big) \big( \delta_{n l^\prime} \, {\cal G}^{l^\prime}_{{\bf k}} (\tau)  + \delta_{n l^\prime +N} \, {\cal G}^{l^\prime}_{-{\bf k}}(-\tau) \big) \nonumber
\end{eqnarray} 
with $1\leq l, l^\prime \leq N$. Thus, the response function $Q_{\mu \nu}(i\omega_m) = - \frac{1}{V}\int_0^{1/T}\langle T_\tau J^z_{s,\mu} (\tau) J_{th, \nu}(0) \rangle e^{i\omega_m \tau} d\tau$ reads
\begin{eqnarray}\label{eq:kernel}
Q_{\mu \nu}(i\omega_m) &=& - \frac{T}{V} \sum_{\varepsilon_n}\sum_{\bf k} \sum_{l, l^\prime=1}^N  \bigg[ \big( \tilde{V}^s_{{\bf k},\mu} \big)_{ll^\prime} \Big\{ \big(  \tilde{V}^{th}_{{\bf k},\nu} \big)_{l^\prime l} +  \big(  \tilde{V}^{th}_{-{\bf k},\nu} \big)_{l+N l^\prime+N} \Big\}  {\cal G}^l_{\bf k}(i\varepsilon_n) {\cal G}^{l^\prime}_{\bf k} (i\varepsilon_n+i\omega_m) \nonumber\\
&& + \big( \tilde{V}^s_{{\bf k},\mu} \big)_{ll^\prime+N} \Big\{ \big( \tilde{V}^{th}_{{\bf k},\nu} \big)_{l^\prime+N l} +  \big(  \tilde{V}^{th}_{-{\bf k},\nu}  \big)_{l+N l^\prime} \Big\}  {\cal G}^l_{\bf k}(i\varepsilon_n) {\cal G}^{l^\prime}_{-{\bf k}} (-i\varepsilon_n-i\omega_m) \nonumber\\
&& + \big(  \tilde{V}^s_{{\bf k},\mu}  \big)_{l+Nl^\prime} \Big\{ \big( \tilde{V}^{th}_{{\bf k},\nu} \big)_{l^\prime l+N} +  \big( \tilde{V}^{th}_{-{\bf k},\nu}  \big)_{l l^\prime+N} \Big\}  {\cal G}^{l}_{-{\bf k}}(i\varepsilon_n) {\cal G}^{l^\prime}_{\bf k} (-i\varepsilon_n+i\omega_m) \nonumber\\
&& + \big( \tilde{V}^s_{{\bf k},\mu} \big)_{l+Nl^\prime+N} \Big\{ \big( \tilde{V}^{th}_{{\bf k},\nu} \big)_{l^\prime+N l+N} +  \big( \tilde{V}^{th}_{-{\bf k},\nu} \big)_{l l^\prime} \Big\}  {\cal G}^{l}_{-{\bf k}}(i\varepsilon_n) {\cal G}^{l^\prime}_{-{\bf k}} (i\varepsilon_n-i\omega_m) \bigg].
\end{eqnarray}
By using the formula $T\sum_{\varepsilon_n} {\cal G}^l_{\bf k}(i\varepsilon_n) {\cal G}^{l^\prime}_{\bf k} (i\varepsilon_n+i\omega_m) = \oint \frac{dz}{2\pi i} f_B(z) G^{l \, R,A}_{\bf k}(z) G^{l^\prime \, R,A}_{\bf k} (z+i\omega_m)$
with the retarded and advanced Green's functions $G^{l \, R}_{\bf k}(z) = [z-\epsilon_{l,{\bf k}} + i\gamma]^{-1}$ and $G^{l \, A}_{\bf k}(z) = [z-\epsilon_{l,{\bf k}} - i\gamma]^{-1}$, one can rewrite the summation over the Matsubara frequency for $\omega_m>0$ as
\begin{eqnarray}\label{eq:Matsubara}
T\sum_{\varepsilon_n} {\cal G}^l_{\bf k}(i\varepsilon_n) {\cal G}^{l^\prime}_{\bf k} (i\varepsilon_n+i\omega_m) &=& \frac{1}{\pi} \int_{-\infty}^\infty dx f_B(x) q_1(x,i\omega_m), \nonumber\\
q_1(x,i\omega_m) &=& G^{l^\prime \, R}_{\bf k}(x+i\omega_m) {\rm Im} G^{l \, R}_{\bf k}(x) + G^{l \, A}_{\bf k}(x-i\omega_m) {\rm Im} G^{l^\prime \, R}_{\bf k}(x) , \\
T\sum_{\varepsilon_n} {\cal G}^l_{\bf k}(i\varepsilon_n) {\cal G}^{l^\prime}_{-{\bf k}} (-i\varepsilon_n-i\omega_m) &=&  \frac{1}{\pi}\int_{-\infty}^\infty dx f_B(x) q_2(x,i\omega_m), \nonumber\\
q_2(x,i\omega_m) &=&  G^{l^\prime \, A}_{-{\bf k}}(-x-i\omega_m) {\rm Im} G^{l \, R}_{\bf k}(x) - G^{l \, A}_{\bf k}(x-i\omega_m) {\rm Im} G^{l^\prime \, R}_{-{\bf k}}(-x) , \nonumber\\
T\sum_{\varepsilon_n} {\cal G}^{l}_{-{\bf k}}(i\varepsilon_n) {\cal G}^{l^\prime}_{\bf k} (-i\varepsilon_n+i\omega_m) &=& \frac{1}{\pi}\int_{-\infty}^\infty dx f_B(x) q_3(x,i\omega_m),\nonumber\\
q_3(x,i\omega_m) &=& - G^{l \, R}_{-{\bf k}}(x+i\omega_m) {\rm Im} G^{l^\prime \, R}_{\bf k}(-x) + G^{l^\prime \, R}_{\bf k}(-x+i\omega_m) {\rm Im} G^{l \, R}_{-{\bf k}}(x),  \nonumber\\
T\sum_{\varepsilon_n} {\cal G}^{l}_{-{\bf k}}(i\varepsilon_n) {\cal G}^{l^\prime}_{-{\bf k}} (i\varepsilon_n-i\omega_m) &=& \frac{1}{\pi}\int_{-\infty}^\infty dx f_B(x) q_4(x,i\omega_m) , \nonumber\\
q_4(x,i\omega_m) &=&  G^{l \, R}_{-{\bf k}}(x+i\omega_m) {\rm Im} G^{l^\prime \, R}_{-{\bf k}}(x) + G^{l^\prime \, A}_{-{\bf k}}(x-i\omega_m) {\rm Im} G^{l \, R}_{-{\bf k}}(x)  . \nonumber
\end{eqnarray}
In the calculation of the spin conductivity $\chi^{\rm SC}_{\mu \nu} = \frac{i}{T} \frac{d Q_{\mu\nu}(\omega + i \gamma )}{d \omega}|_{\omega=0}$, the derivative with respect to $\omega$ acts only on $q_i$ in Eq. (\ref{eq:Matsubara}), and we have 
\begin{eqnarray}\label{eq:Matsubara_div}
\frac{\partial q_1(x,\omega)}{\partial \omega} \Big|_{\omega=0} &=& \frac{\gamma}{(x-\epsilon_{l,{\bf k}})^2+\gamma^2}\frac{(x-\epsilon_{l^\prime, {\bf k}}-i\gamma)^2}{[(x-\epsilon_{l^\prime,{\bf k}})^2+\gamma^2]^2} - \frac{\gamma}{(x-\epsilon_{l^\prime,{\bf k}})^2+\gamma^2}\frac{(x-\epsilon_{l, {\bf k}}+i\gamma)^2}{[(x-\epsilon_{l,{\bf k}})^2+\gamma^2]^2} , \\
\frac{\partial q_2(x,\omega)}{\partial \omega} \Big|_{\omega=0}&=& -\frac{\gamma}{(x-\epsilon_{l,{\bf k}})^2+\gamma^2}\frac{(x+\epsilon_{l^\prime, -{\bf k}}-i\gamma)^2}{[(x+\epsilon_{l^\prime,-{\bf k}})^2+\gamma^2]^2} + \frac{\gamma}{(x+\epsilon_{l^\prime,-{\bf k}})^2+\gamma^2}\frac{(x-\epsilon_{l, {\bf k}}+i\gamma)^2}{[(x-\epsilon_{l,{\bf k}})^2+\gamma^2]^2} , \nonumber\\
\frac{\partial q_3(x,\omega)}{\partial \omega} \Big|_{\omega=0} &=& -\frac{\gamma}{(x+\epsilon_{l^\prime,{\bf k}})^2+\gamma^2}\frac{(x-\epsilon_{l, -{\bf k}}-i\gamma)^2}{[(x-\epsilon_{l,-{\bf k}})^2+\gamma^2]^2} + \frac{\gamma}{(x-\epsilon_{l,-{\bf k}})^2+\gamma^2}\frac{(x+\epsilon_{l^\prime, {\bf k}}+i\gamma)^2}{[(x+\epsilon_{l^\prime,{\bf k}})^2+\gamma^2]^2} , \nonumber\\
\frac{\partial q_4(x,\omega)}{\partial \omega} \Big|_{\omega=0}&=& \frac{\gamma}{(x-\epsilon_{l^\prime,-{\bf k}})^2+\gamma^2}\frac{(x-\epsilon_{l, -{\bf k}}-i\gamma)^2}{[(x-\epsilon_{l,-{\bf k}})^2+\gamma^2]^2} - \frac{\gamma}{(x-\epsilon_{l,-{\bf k}})^2+\gamma^2}\frac{(x-\epsilon_{l^\prime, -{\bf k}}+i\gamma)^2}{[(x-\epsilon_{l^\prime,-{\bf k}})^2+\gamma^2]^2} . \nonumber
\end{eqnarray} 
The $\gamma \rightarrow 0$ limit corresponds to the free magnon without a damping. As our interest here is in the longitudinal response to the thermal gradient, magnon scatterings should be nonnegligible as in the case of the electric conductivity. In this work, we introduce the magnon damping in the form of a nonzero but small constant $\gamma$, mimicking the derivation of the electric conductivity in the Kubo-formula approach \cite{book_Mahan}. 
This corresponds to a relaxation-time approximation where the magnon scattering time $\tau_m$, which is related to the damping via $\gamma=1/\tau_m$, is assumed to be a constant. In reality, the magnon damping depends on the wave number, energy, and temperature due to, for example, magnon-magnon and magnon-phonon scatterings \cite{Magdamp_Cornelissen_prb_16}. As our focus here is on the qualitative understanding of the role of the magnon band for the magnon transport, we use this simplified assumption of the constant $\gamma$, ignoring the complicated parameter dependence of the magnon damping.

For an energy scale $x$ satisfying $\gamma \ll |x-\epsilon_{l,{\bf k}}|$, we use the approximations \cite{trans-sq_AK_prb_19, trans-tri_AK_prl_20, trans_Tatara_prb_15}
\begin{equation}
\frac{\gamma}{(x-\epsilon_{l,{\bf k}})^2+\gamma^2} \simeq \pi \delta(x-\epsilon_{l,{\bf k}}), \quad \frac{\gamma^3}{[(x-\epsilon_{l,{\bf k}})^2+\gamma^2]^2} \simeq \frac{\pi}{2} \delta(x-\epsilon_{l,{\bf k}})
\end{equation}
as functions of the variable $x$. Then, for the contribution from $q_1$, by performing the following integral by parts, we obtain
\begin{eqnarray}\label{eq:integral}
&& \frac{1}{\pi} \int_{-\infty}^\infty dx f_B(x)\frac{\partial q_1(x,\omega)}{\partial \omega} \Big|_{\omega=0} = -i\gamma^2 \frac{1}{\pi}\int_{-\infty}^\infty dx f^\prime_B(x)\frac{1}{[(x-\epsilon_{l,{\bf k}})^2+\gamma^2][(x-\epsilon_{l^\prime,{\bf k}})^2+\gamma^2]} \nonumber\\
&&\qquad + 2\gamma^3 \frac{1}{\pi} \int_{-\infty}^\infty dx f_B(x)\Big\{ \frac{1}{[(x-\epsilon_{l,{\bf k}})^2+\gamma^2]^2[(x-\epsilon_{l^\prime,{\bf k}})^2+\gamma^2]} - \frac{1}{[(x-\epsilon_{l,{\bf k}})^2+\gamma^2][(x-\epsilon_{l^\prime,{\bf k}})^2+\gamma^2]^2} \Big\} \nonumber\\
 &&\qquad \simeq \left\{ \begin{array}{l}
\displaystyle{  -i \frac{1}{2 \gamma} \, f^\prime_B(\epsilon_{l,{\bf k}}) \qquad \qquad (\epsilon_{l,{\bf k}}=\epsilon_{l^\prime,{\bf k}})  } \\
\displaystyle{  -i\gamma \frac{f^\prime_B(\epsilon_{l,{\bf k}})+f^\prime_B(\epsilon_{l^\prime,{\bf k}}) }{(\epsilon_{l,{\bf k}}-\epsilon_{l^\prime,{\bf k}})^2} + \frac{f_B(\epsilon_{l,{\bf k}})-f_B(\epsilon_{l^\prime,{\bf k}}) }{(\epsilon_{l,{\bf k}}-\epsilon_{l^\prime,{\bf k}})^2}  -4 \gamma^2\frac{f_B(\epsilon_{l,{\bf k}})-f_B(\epsilon_{l^\prime,{\bf k}}) }{(\epsilon_{l,{\bf k}}-\epsilon_{l^\prime,{\bf k}})^4}  } \quad (\epsilon_{l,{\bf k}} \neq \epsilon_{l^\prime,{\bf k}})
\end{array} \right. .
\end{eqnarray} 
It turns out that the $1/\gamma$ term in Eq. (\ref{eq:integral}) yields a contribution proportional to $1/\gamma$ in the conductivity $\chi^{\rm SC}_{\mu \nu} = \frac{i}{T} \frac{d Q_{\mu\nu}(\omega + i \gamma )}{d \omega}|_{\omega=0}$, as is expected in the Boltzmann approximation. In the same  manner as that for Eq. (\ref{eq:integral}), one can evaluate the contribution from $q_2$ as
\begin{equation}
\frac{1}{\pi} \int_{-\infty}^\infty dx f_B(x)\frac{\partial q_2(x,\omega)}{\partial \omega} \Big|_{\omega=0} \simeq   i\gamma \frac{f^\prime_B(\epsilon_{l,{\bf k}})+f^\prime_B(-\epsilon_{l^\prime,-{\bf k}}) }{(\epsilon_{l,{\bf k}}+\epsilon_{l^\prime,-{\bf k}})^2} + \frac{f_B(\epsilon_{l,{\bf k}})-f_B(-\epsilon_{l^\prime,-{\bf k}}) }{(\epsilon_{l,{\bf k}}+\epsilon_{l^\prime,-{\bf k}})^2}  +4 \gamma^2\frac{f_B(\epsilon_{l,{\bf k}})-f_B(-\epsilon_{l^\prime,-{\bf k}}) }{(\epsilon_{l,{\bf k}}+\epsilon_{l^\prime,-{\bf k}})^4}  \nonumber
\end{equation} 
In contrast to the $q_1$ term, the $q_2$ term does not contain a term proportional to $1/\gamma$. 
By picking up the terms proportional to $1/\gamma$ in $\int_{-\infty}^\infty dx f_B(x)\frac{\partial q_i(x,\omega)}{\partial \omega} |_{\omega=0}$ $(i=1-4)$, we obtain
\begin{eqnarray}
\chi^{\rm SC}_{\mu \nu} &=&  -\frac{1}{VT}\frac{1}{2\gamma} \sum_{\bf k} \sum_{l,l^\prime =1}^N \delta_{\epsilon_{l,{\bf k}} \epsilon_{l^\prime,{\bf k}}} \bigg[  f^\prime_B(\epsilon_{l,{\bf k}})  \big( \tilde{V}^s_{{\bf k},\mu} \big)_{ll^\prime} \Big\{ \big(  \tilde{V}^{th}_{{\bf k},\nu} \big)_{l^\prime l} +  \big(  \tilde{V}^{th}_{-{\bf k},\nu} \big)_{l+N l^\prime+N} \Big\} \nonumber\\
&&\qquad\qquad\qquad\qquad\qquad + f^\prime_B(\epsilon_{l,-{\bf k}}) \big( \tilde{V}^s_{{\bf k},\mu} \big)_{l+Nl^\prime+N} \Big\{ \big( \tilde{V}^{th}_{{\bf k},\nu} \big)_{l^\prime+N l+N} +  \big( \tilde{V}^{th}_{-{\bf k},\nu} \big)_{l l^\prime} \Big\} \bigg].
\end{eqnarray}
Since $\epsilon_{l,{\bf k}} = \epsilon_{l^\prime,{\bf k}}$ is satisfied over the whole momentum space for $l=l^\prime$ and only at degeneracy points for $l \neq l^\prime$, the dominant contribution comes from the former, so that we could write
\begin{equation}
\chi^{\rm SC}_{\mu \nu} =  -\frac{1}{VT}\frac{1}{2\gamma} \sum_{\bf k} \sum_{l =1}^N  f^\prime_B(\epsilon_{l,{\bf k}}) \bigg[  \big( \tilde{V}^s_{{\bf k},\mu} \big)_{ll} \Big\{ \big(  \tilde{V}^{th}_{{\bf k},\nu} \big)_{l l} +  \big(  \tilde{V}^{th}_{-{\bf k},\nu} \big)_{l+N l+N} \Big\} + \big( \tilde{V}^s_{{\bf k},\mu} \big)_{l+Nl+N} \Big\{ \big( \tilde{V}^{th}_{{\bf k},\nu} \big)_{l+N l+N} +  \big( \tilde{V}^{th}_{-{\bf k},\nu} \big)_{l l} \Big\} \bigg],
\end{equation}
where we have used the relation $\epsilon_{l, {\bf k}}=\epsilon_{l,-{\bf k}}$ satisfied in the present system [see Fig. 2 (c) in the main text]. We note in passing that for noncoplanar spin structures, the magnon dispersion can become asymmetric $\epsilon_{l, {\bf k}}\neq \epsilon_{l,-{\bf k}}$ due to spatially inhomogeneous geometric phases \cite{KA_BrKagome_SW_prb_25}. 
By further using the relation $(T_{-{\bf k}})_{\alpha \beta}= (\sigma_1 T^\ast_{\bf k} \sigma_1)_{\alpha \beta} \exp[i\theta_{{\bf k},\beta}]$ \cite{LinearRes_Matsumoto_prb_14}, one can rewrite $(\tilde{V}^{th}_{-{\bf k},\mu})_{ll}$ and $(\tilde{V}^{th}_{-{\bf k},\mu})_{l+Nl+N}$ into $(\tilde{V}^{th}_{{\bf k},\mu})_{l+Nl+N}$ and $(\tilde{V}^{th}_{{\bf k},\mu})_{ll}$, respectively. Thus, we finally obtain
\begin{equation}\label{eq:response_Boltz}
\chi^{\rm SC}_{\mu \nu} =  -\frac{1}{VT}\frac{1}{\gamma} \sum_{\bf k} \sum_{l =1}^N  f^\prime_B(\epsilon_{l,{\bf k}}) \bigg[  \big( \tilde{V}^s_{{\bf k},\mu} \big)_{ll} \big(  \tilde{V}^{th}_{{\bf k},\nu} \big)_{l l}  + \big( \tilde{V}^s_{{\bf k},\mu} \big)_{l+Nl+N} \big( \tilde{V}^{th}_{{\bf k},\nu} \big)_{l+N l+N} \bigg].
\end{equation}
Since both $\tilde{V}^s_{{\bf k},\mu}$ and $\tilde{V}^{th}_{{\bf k}, \mu}$ are linear in the magnon velocity $v_{{\bf k},\mu} = \frac{\partial H_{\bf k}}{\partial q_\mu}$, $\chi^{\rm SC}_{\mu\nu}$ in Eq. (\ref{eq:response_Boltz}) takes the form of $\chi^{\rm SC}_{\mu\nu} \sim \frac{1}{\gamma} \sum_{\bf k} \, f^\prime_B(\epsilon_{\bf k})\, v_{{\bf k},\mu}  v_{{\bf k},\nu} $, which is definitely in the same form as that in the Boltzmann approximation. In Eq. (\ref{eq:response_Boltz}), the summation over ${\bf k}$ can numerically be evaluated by using the replacement $\frac{1}{V}\sum_{\bf k} \rightarrow \frac{1}{(2\pi)^2}\int_{\rm MBZ} d{\bf k}$, where ${\rm MBZ}$ denotes the magnetic Brillouin zone. 

 In the case of the thermal conductivity $\kappa_{\mu \nu} = \frac{i}{T} \frac{d Q^{th}_{\mu\nu}(\omega + i \gamma )}{d \omega}|_{\omega=0}$ with $Q^{th}_{\mu \nu}(i\omega_m) = - \frac{1}{V}\int_0^{1/T}\langle T_\tau J_{th,\mu} (\tau) J_{th, \nu}(0) \rangle e^{i\omega_m \tau} d\tau$, we only replace $\tilde{V}^s_{{\bf k}, \mu}$ with $\tilde{V}^{th}_{{\bf k}, \mu}$ in Eq. (\ref{eq:response_Boltz}). 



\begin{thebibliography}{80}

\bibitem{alter_Smejkal_prx_22_overview} L. \v{S}mejkal, J. Sinova, and T. Jungwirth, Emerging Research Landscape of Altermagnetism, Phys. Rev. X {\bf 12}, 040501 (2022). 
\bibitem{alter_Noda_pccp_16} Y. Noda, K. Ohno, and S. Nakamura, Momentum-dependent band spin splitting in semiconducting MnO$_2$: a density functional calculation, Phys. Chem. Chem. Phys. {\bf 18}, 13294 (2016).
\bibitem{alter_Naka_natcom_19} M. Naka, S. Hayami, H. Kusunose, Y. Yanagi, Y. Motome, and H. Seo, Spin current generation in organic antiferromagnets, Nat. Commun. {\bf 10}, 4305 (2019). 
\bibitem{alter_Hayami_jpsj_19} S. Hayami, Y. Yanagi, and H. Kusunose, Momentum-Dependent Spin Splitting by Collinear Antiferromagnetic Ordering, J. Phys. Soc. Jpn. {\bf 88}, 123702 (2019).  
\bibitem{alter_Yuan_prb_20} L.-D. Yuan, Z. Wang, J.-W. Luo, E. I. Rashba, and A. Zunger, Giant momentum-dependent spin splitting in centrosymmetric low-$Z$ antiferromagnets, Phys. Rev. B {\bf 102}, 014422 (2020). 


\bibitem{cferri_Groot_prl_95} H. van Leuken and R. A. de Groot, Half-Metallic Antiferromagnets, Phys. Rev. Lett. {\bf 74}, 1171 (1995).
\bibitem{cferri_Pickett} W. E. Pickett, Spin-density-functional-based search for half-metallic antiferromagnets, Phys. Rev. B {\bf 57}, 10613 (1998).
\bibitem{cferri_Misawa_prl_24} T. Kawamura, K. Yoshimi, K. Hashimoto, A. Kobayashi, and T. Misawa, Compensated Ferrimagnets with Colossal Spin Splitting in Organic Compounds, Phys. Rev. Lett. {\bf 132}, 156502 (2024).
\bibitem{cferri_Liu_prl_25} Y. Liu, S.-D. Guo, Y. Li, and C.-C. Liu, Two-Dimensional Fully Compensated Ferrimagnetism, Phys. Rev. Lett. {\bf 134}, 116703 (2025).
\bibitem{cferri_Lee_arX_25} X. T. Lee, T. Misawa, M. Matsuo, and T. Kato, Spin transport phenomena in junctions composed of a compensated ferrimagnet and a normal metal, arXiv:2507.05618.

\bibitem{alter_Mazin_review_24} I. Mazin, Editorial: Altermagnetism - A New Punch Line of Fundamental Magnetism, Phys. Rev. X {\bf 12}, 040002 (2022).


\bibitem{alter_Hernandez_prl_21} R. G.-Hern\'{a}ndez, L. \v{S}mejkal, K. V\'{y}born\'{y}, Y. Yahagi, J. Sinova, T. Jungwirth, and J. \v{Z}elezn\'{y}, Efficient Electrical Spin Splitter Based on Nonrelativistic Collinear Antiferromagnetism, Phys. Rev. Lett. {\bf 26}, 127701 (2021). 
\bibitem{alter_Smejkal_prx_22} L. \v{S}mejkal, J. Sinova, and T. Jungwirth, Beyond Conventional Ferromagnetism and Antiferromagnetism: A Phase with Nonrelativistic Spin and Crystal Rotation Symmetry, Phys. Rev. X {\bf 12}, 031042 (2022). 
\bibitem{alter_Yuan_prb_21} L.-D. Yuan, Z. Wang, J.-W. Luo, and A. Zunger, Prediction of low-$Z$  collinear and noncollinear antiferromagnetic compounds having momentum-dependent spin splitting even without spin-orbit coupling, Phys. Rev. Mater. {\bf 5}, 014409 (2021). 


\bibitem{fcc_Henly_87} C. L. Henly, Ordering by disorder: Ground-state selection in fcc vector antiferromagnets, J. Appl. Phys. {\bf 61}, 3962 (1987).
\bibitem{tri_Kawamura_85} H. Kawamura and S. Miyashita, Phase Transition of the Heisenberg Antiferromagnet on the Triangular Lattice in a Magnetic Field, J. Phys. Soc. Jpn. {\bf 85}, 4530 (1985).
\bibitem{tri_Chubkov_91} A. V. Chubukov and D. I. Golosov, Quantum theory of an antiferromagnet on a triangular lattice in a magnetic field, J. Phys.: Condens. Matter {\bf 3}, 69 (1991).


\bibitem{RMO-collinear_Grison_prb_20} V. Grison, P. Viot, B. Bernu, and L. Messio, Emergent Potts order in the kagome $J_1$-$J_3$ Heisenberg model, Phys. Rev. B {\bf 102}, 214424 (2020).
\bibitem{KagomeSkX_AK_22} K. Aoyama and H. Kawamura, Emergent skyrmion-based chiral order in zero-field Heisenberg antiferromagnets on the breathing kagome lattice, Phys. Rev. B {\bf 105}, L100407 (2022). 
\bibitem{KagomeSkX_AK_23} K. Aoyama and H. Kawamura, Zero-Field Miniature Skyrmion Crystal and Chiral Domain State in Breathing-Kagome Antiferromagnets, J. Phys. Soc. Jpn. {\bf 92}, 033701 (2023). 
 

\bibitem{CoplanarOct_Boldrin_prl_18} D. Boldrin, B. Fak, E. Can\'{e}vet, J. Ollivier, H. C. Walker, P. Manuel, D. D. Khalyavin, and A. S. Wills, Vesignieite: An $S=\frac{1}{2}$ Kagome Antiferromagnet with Dominant Third-Neighbor Exchange, Phys. Rev. Lett. {\bf 121}, 107203 (2018).


\bibitem{LinearRes_Matsumoto_prb_14} R. Matsumoto, R. Shindou, and S. Murakami, Thermal Hall effect of magnons in magnets with dipolar interaction, Phys. Rev. B {\bf 89}, 054420 (2014).



\bibitem{alter_magnon_Smejkal_prl_23} L. \v{S}mejkal, A. Marmodoro, K.-H. Ahn, R. G.-H\'{e}rnandez, I. Turek, S. Mankovsky, H. Ebert, S. W. D\'Souza, O. \v{S}ipr, J. Sinova, and T. Jungwirth, Chiral Magnons in Altermagnetic RuO$_2$, Phys. Rev. Lett. {\bf 131}, 256703 (2023). 
\bibitem{alter_magnon_Liu_prl_24} Z. Liu, M. Ozeki, S. Asai, S. Itoh, and T. Masuda, Chiral Split Magnon in Altermagnetic MnTe, Phys. Rev. Lett. {\bf 133}, 156702 (2024).


\bibitem{MagnonPol_Nambu_prl_20} Y. Nambu, J. Barker, Y. Okino, T. Kikkawa, Y. Shiomi, M. Enderle, T. Weber, B. Winn, M. Graves-Brook, J. M. Tranquada, T. Ziman, M. Fujita, G. E.W. Bauer, E. Saitoh, and K. Kakurai, Observation of Magnon Polarization, Phys. Rev. Lett. {\bf 125}, 027201 (2020).
\bibitem{MagnonPol_Kawamoto_apl_24} Y. Kawamoto, T. Kikkawa, M. Kawamata, Y. Umemoto, A. G. Manning, K. C. Rule, K. Ikeuchi, K. Kamazawa, M. Fujita, E. Saitoh, K. Kakurai, and Y. Nambu, Understanding spin currents from magnon dispersion and polarization: Spin-Seebeck effect and neutron scattering study on Tb$_3$Fe$_5$O$_{12}$, App. Phys. Lett. {\bf 124}, 132406 (2024).


\bibitem{alter_qfluc_Eto_prb_25} R. Eto, M. Gohlke, J. Sinova, M. Mochizuki, A. L. Chernyshev, and A. Mook, Spontaneous magnon decays from nonrelativistic time-reversal symmetry breaking in altermagnets, Phys. Rev. B {\bf 112}, 094442 (2025).
\bibitem{alter_qfluc_Cichutek_prb_25} N. Cichutek, P. Kopietz, and A. R\"{u}ckriegel, Quantum fluctuations in two-dimensional altermagnets, Phys. Rev. B {\bf 112}, 094442 (2025).


\bibitem{book_Mahan} G. D. Mahan, {\it Many-Particle Physics}, third edition (Springer Science+Business Media, New York, 2000).
\bibitem{trans-sq_AK_prb_19} K. Aoyama and H. Kawamura, Effects of magnetic anisotropy on spin and thermal transports in classical antiferromagnets on the square lattice, Phys. Rev. B {\bf 100}, 144416 (2019).
\bibitem{trans-tri_AK_prl_20}  K. Aoyama and H. Kawamura, Spin Current as a Probe of the $\mathbb{Z}_2$-Vortex Topological Transition in the Classical Heisenberg Antiferromagnet on the Triangular Lattice, Phys. Rev. Lett. {\bf 124}, 047202 (2020).


\bibitem{supple} K. Aoyama and H. Kawamura, Supplemental Material which includes Refs. \cite{LinearRes_Matsumoto_prb_14, book_Mahan, Magdamp_Cornelissen_prb_16,  trans-sq_AK_prb_19, trans-tri_AK_prl_20, trans_Tatara_prb_15, KA_BrKagome_SW_prb_25}, where the derivation of the conductivities $\chi^{\rm SC}_{\mu \nu}$ and $\kappa_{\mu\nu}$ is presented.

\bibitem{Magdamp_Cornelissen_prb_16} L. J. Cornelissen, K. J. H. Peters, G. E. W. Bauer, R. A. Duine, and B. J. van Wees, Magnon spin transport driven by the magnon chemical potential in a magnetic insulator, Phys. Rev. B {\bf 94}, 014412 (2016).


\bibitem{trans_Tatara_prb_15} G. Tatara, Thermal vector potential theory of magnon-driven magnetization dynamics, Phys. Rev. B {\bf 92}, 064405 (2015).
\bibitem{KA_BrKagome_SW_prb_25} K. Aoyama and H. Kawamura, Spontaneous chirality selection and nonreciprocal spin wave in breathing-kagome antiferromagnets at zero field, Phys. Rev. B {\bf 111}, 144413 (2025).

\bibitem{cferi_Seebeck_Maekawa_prb_13} Y. Ohnuma, H. Adachi, E. Saitoh, and S. Maekawa, Spin Seebeck effect in antiferromagnets and compensated ferrimagnets, Phys. Rev. B {\bf 87}, 014423 (2013).

\cite{CoplanarOct_Boldrin_prl_18}
\bibitem{Sr-vesignieite_Boldrin_jmc_15} D. Boldrin and A. S. Wills, SrCu$_3$V$_2$O$_8$(OH)$_2$ – dynamic Jahn–Teller distortions and orbital frustration in a new $S=1/2$ kagome antiferromagnet, J. Mater. Chem. C {\bf 3}, 4308 (2015).
\bibitem{Ba-vesignieite_Boldrin_jmc_16} D. Boldrin, K. Knight, and A. S. Wills, Orbital frustration in the $S =1/2$ kagome magnet vesignieite, BaCu$_3$V$_2$O$_8$(OH)$_2$, J. Mater. Chem. C {\bf 4}, 10315 (2016).
\bibitem{Sr-vesignieite_DM_Verrier_prb_20} A. Verrier, F. Bert, J. M. Parent, M. El-Amine, J. C. Orain, D. Boldrin, A. S. Wills, P. Mendels, and J. A. Quilliam, Canted antiferromagnetic order in the kagome material Sr-vesignieite, Phys. Rev. B {\bf 101}, 054425 (2020).

\bibitem{Artificial-kagome_Zhou_natcom_24} H. Zhou, M. S. Dias, Y. Zhang, W. Zhao, and S. Lounis, Kagomerization of transition metal monolayers induced by two-dimensional hexagonal boron nitride, Nat. Commun. {\bf 15}, 4854 (2024).


\bibitem{orbital_alter_Mook_prl_24} V. Leeb, A. Mook, L. \v{S}mejkal, and J. Knolle, Spontaneous Formation of Altermagnetism from Orbital Ordering, Phys. Rev. Lett. {\bf 132}, 236701 (2024).
 
\bibitem{Flat_Mizoguchi_prb_19} T. Mizoguchi and M. Udagawa, Flat-band engineering in tight-binding models: Beyond the nearest-neighbor hopping, Phys. Rev. B {\bf 99}, 235118 (2019).

\end{thebibliography}

\begin{thebibliography}{80}
\bibitem{book_Mahan} G. D. Mahan, {\it Many-Particle Physics}, third edition (Springer Science+Business Media, New York, 2000).
\bibitem{Magdamp_Cornelissen_prb_16} L. J. Cornelissen, K. J. H. Peters, G. E. W. Bauer, R. A. Duine, and B. J. van Wees, Magnon spin transport driven by the magnon chemical potential in a magnetic insulator, Phys. Rev. B {\bf 94}, 014412 (2016).
\bibitem{trans-sq_AK_prb_19} K. Aoyama and H. Kawamura, Effects of magnetic anisotropy on spin and thermal transports in classical antiferromagnets on the square lattice, Phys. Rev. B {\bf 100}, 144416 (2019).
\bibitem{trans-tri_AK_prl_20}  K. Aoyama and H. Kawamura, Spin Current as a Probe of the $\mathbb{Z}_2$-Vortex Topological Transition in the Classical Heisenberg Antiferromagnet on the Triangular Lattice, Phys. Rev. Lett. {\bf 124}, 047202 (2020).
\bibitem{trans_Tatara_prb_15} G. Tatara, Thermal vector potential theory of magnon-driven magnetization dynamics, Phys. Rev. B {\bf 92}, 064405 (2015).
\bibitem{KA_BrKagome_SW_prb_25} K. Aoyama and H. Kawamura, Spontaneous chirality selection and nonreciprocal spin wave in breathing-kagome antiferromagnets at zero field, Phys. Rev. B {\bf 111}, 144413 (2025).
\bibitem{LinearRes_Matsumoto_prb_14} R. Matsumoto, R. Shindou, and S. Murakami, Thermal Hall effect of magnons in magnets with dipolar interaction, Phys. Rev. B {\bf 89}, 054420 (2014).

\end{thebibliography}
\end{document}